\documentclass[sigconf]{acmart}

\AtBeginDocument{%
  \providecommand\BibTeX{{%
    \normalfont B\kern-0.5em{\scshape i\kern-0.25em b}\kern-0.8em\TeX}}}

\usepackage{multirow}
\usepackage{booktabs}
\usepackage{cellspace}
\usepackage{hhline} 
\usepackage{graphicx}
\usepackage{xcolor}
\usepackage{diagbox}
\usepackage{arydshln}

\copyrightyear{2024} 
\acmYear{2024} 
\setcopyright{rightsretained} 
\acmConference[CHI EA '24]{Extended Abstracts of the CHI Conference on
Human Factors in Computing Systems}{May 11--16, 2024}{Honolulu, HI, USA}
\acmBooktitle{Extended Abstracts of the CHI Conference on Human Factors in
Computing Systems (CHI EA '24), May 11--16, 2024, Honolulu, HI, USA}
\acmDOI{10.1145/3613905.3650952}
\acmISBN{979-8-4007-0331-7/24/05}

\begin{document}

\title{CanvasPic: An Interactive Tool for Freely Generating Facial Images Based on Spatial Layout}

\author{Jiafu Wei}
\email{weijf21@mails.jlu.edu.cn}
\affiliation{%
  \institution{Jilin University}
  \city{Chang Chun}
  \country{China}}

\author{Chia-Ming Chang}
\email{info@chiamingchang.com}
\affiliation{%
  \institution{The University of Tokyo}
  \city{Tokyo}
  \country{Japan}}

\author{Xi Yang}
\authornote{Corresponding author. Engineering Research Center of Knowledge-Driven Human-Machine Intelligence, MoE. Jilin Province Key Laboratory of Ancient Chinese Script, Culture relics and Artificial Intelligence.}
\email{yangxi21@jlu.edu.cn}
\affiliation{%
  \institution{Jilin University}
  \city{Chang Chun}
  \country{China}}

\author{Takeo Igarashi}
\email{takeo@acm.org}
\affiliation{%
  \institution{The University of Tokyo}
  \city{Tokyo}
  \country{Japan}}

\begin{abstract}
  In real-world usage, existing GAN image generation tools come up short due to their lack of intuitive interfaces and limited flexibility. To overcome these limitations, we developed CanvasPic, an innovative tool for flexible GAN image generation. Our tool introduces a novel 2D layout design that allows users to intuitively control image attributes based on real-world images. By interacting with the distances between images in the spatial layout, users are able to conveniently control the influence of each attribute on the target image and explore a wide range of generated results. Considering practical application scenarios, a user study involving 24 participants was conducted to compare our tool with existing tools in GAN image generation. The results of the study demonstrate that our tool significantly enhances the user experience, enabling more effective achievement of desired generative results.
\end{abstract}




\begin{teaserfigure}
  \centering
  \includegraphics[width=0.95\textwidth]{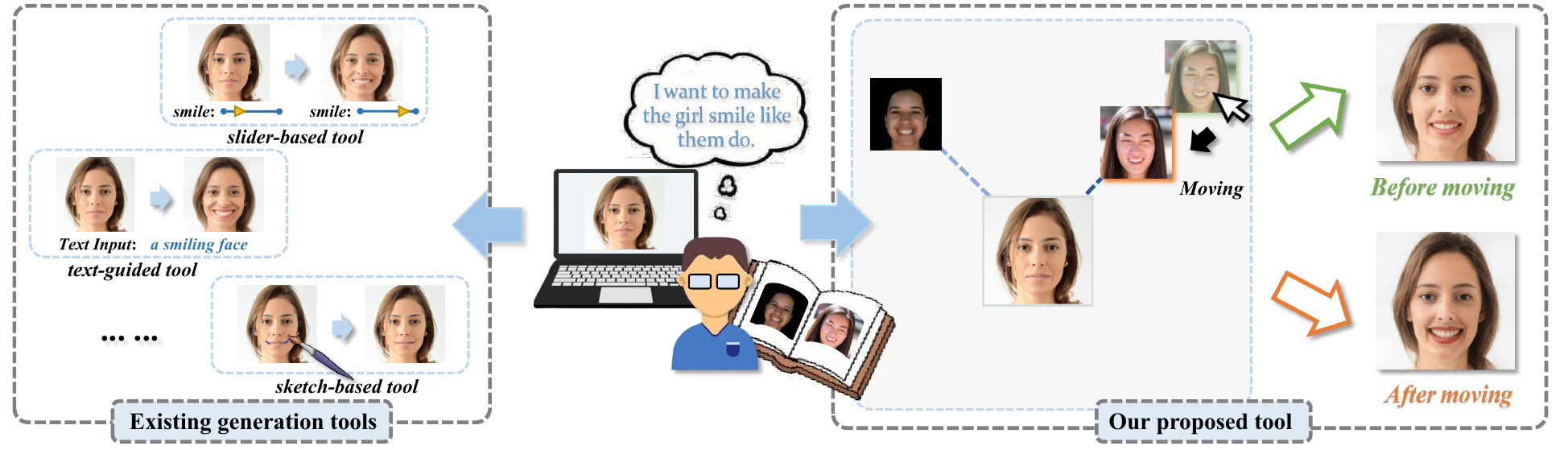}
  \caption{We introduce CanvasPic, a novel generation tool with a 2D spatial layout that allows users to import real-world images as references and generate desired results by adjusting the distances and attributes of these reference images.}
  \label{fig:teaser}
\end{teaserfigure}

\maketitle

\section{Introduction}

Despite the impressive capabilities of Generative Adversarial Networks (GANs), controlling the style and content of their generated images remains a challenge~\cite{rel,rpeng}. Providing active control over GAN-generated image style and content is vital for meeting specific requirements in practical applications. This has led to the emergence of GAN image generation tools that empower users to express their creativity and ideas. Furthermore, personalized generation has gained recognition, resulting in various tools aiming to fulfill users' aspirations. 

However, the existing tools still lack control flexibility and user-friendliness. For example, sketch-based~\cite{r5, r4} tools often demand users to possess specific drawing skills or image editing experience. Additionally, the generated images often do not achieve the desired level of realism. Slider-based tools~\cite{rganspace,rhfgi} offer limited choices, neglecting diverse user needs and creative expressions. Text-based tools~\cite{r1126,r1131,rstyleclip} rely on abstract input, resulting in a gap between the user's expectations and the obtained results. Furthermore, some tools~\cite{rganslider} suffer from intuitive efficiency issues, such as cluttered interfaces and complex operations. As users increasingly seek greater freedom and personalization in generative tools, traditional modes of tools have become inadequate to meet the demands of specific scenarios.

To address the aforementioned challenges, we introduce CanvasPic, a novel tool designed for flexible image generation. The primary goal of CanvasPic is to offer a multifunctional image generation platform that goes beyond traditional control functionalities. Through an innovative 2D layout design, users can effortlessly adjust image positions to control generated results, enhancing both intuitive interaction and efficiency. This method grants users more control over results, elevating their creative freedom. Additionally, CanvasPic integrates real-world images as references, enabling users to guide the generation of target images using attributes from existing images. This not only aids result anticipation but also inspires creative exploration and experimentation, fostering a stronger user-content connection. We generate facial images in this paper which is one of the most general domains, however, our tool is not limited to facial images.

The main contributions of our work are summarized as follows:
\begin{enumerate}   
    \item We developed CanvasPic, a GAN-based tool that empowers users to leverage attributes from real-world images for controlling the generation results of portrait images. 
    \item We introduce a novel 2D spatial layout interface that enables users to control image attributes through distance adjustments. This design fosters user exploration and anticipation of generated results, enhancing task intuitiveness.
    \item We conducted a user study involving 24 participants to evaluate user satisfaction with CanvasPic. The results of the user study demonstrate the significant impact of CanvasPic in enhancing the overall user experience.
\end{enumerate}

\section{Related work}

\subsection{Advancing Image Generation: From GANs to Real-World Synthesis}

In the field of image generation, prominent techniques include the Diffusion Model~\cite{rqdm}, VAE~\cite{rqvae}, and GAN~\cite{rqgan}. Each exhibits distinct strengths: Diffusion Model excels in producing high-quality, realistic natural images; VAE is suitable for image reconstruction and variation generation, while GAN performs well in generating synthetic images closely resembling real ones, particularly in attribute control~\cite{rqzs1,rqzs2}. Leveraging the characteristics of GAN, various image generation technologies based on GAN have been developed, including the influential StyleGAN~\cite{r11stylegan}. For example, researchers~\cite{r113,r114} proposed new generation methods using latent code transformation. Khodadadeh et al.~\cite{r116} achieved successful generation tasks by mapping latent vectors to target images. Liu et al.'s STGAN~\cite{r117} improved attribute manipulation accuracy through selective transfer, while He et al.'s AttGAN~\cite{r11attgan} introduced attribute classification constraints for high-quality generation. However, these methods have difficulty controlling real-world images.

In recent years, research has shifted toward applying image-generation techniques to real-world scenarios. This led to the development of GAN inversion~\cite{r11p2p,r11e4e}, including text-guided~\cite{r1126,r1131} and sketch-based methods~\cite{r5,r4,rsketch}. Especially, Patashnik et al.'s StyleCLIP~\cite{rstyleclip} merged StyleGAN and CLIP models, enabling text-driven image generation based on users' textual descriptions. Ko et al.~\cite{rwetoon} devised a GAN-based communication system for efficient interactions and sketch adjustments among webcomic authors and artists.

Moreover, there has been a proliferation of text-based generative inpainting methods in recent years~\cite{in3,in2,in1}. These approaches utilize textual descriptions to guide the image inpainting process, amalgamating text descriptions with image generation to fully exploit the strengths of both modalities. In contrast to traditional methods solely reliant on textual guidance, integrating local image information enables a more precise comprehension of user intent and requirements, thereby facilitating semantically and contextually enhanced image inpainting.

However, existing image generation tools still suffer from limitations in control flexibility and user-friendliness, making them difficult to effectively utilize in practical scenarios.

\subsection{Enhancing User Experience Through Spatial Layouts}

The concept of spatial layout has broad applications and is often used to define relationships among icons, objects, and information~\cite{r1144,r119,r1110,r1147}. Bauer et al.~\cite{r1141} summarized the concept of a $``$pile$"$ to showcase the benefits of structured spatial layouts. Yi et al.~\cite{rmagnet} introduced the $``$Dust \& Magnet$"$ technique, utilizing magnet metaphor and interactive methods to enhance the understanding and accessibility of multidimensional data. Watanabe et al.~\cite{r1142} applied spatial layout to design an interface for object space aggregation to streamline information management.

Well-designed spatial layouts significantly enhance task efficiency and user experience~\cite{rwei1,rwei2}. For example, Kleiman et al.\cite{r1145} introduced a system enabling users to interact with a large image collection on a 2D canvas based on nearest neighbors in a high-dimensional feature space, resulting in efficient task interaction. Brown et al.\cite{r1143} proposed a system that allows experts to directly interact with visual data representation, ensuring scalability in data dimensions and streamlined task processing. Additionally, Chang et al.~\cite{r11chiaming} designed a spatial layout annotation interface to improve image annotation quality by non-expert annotators.

% Therefore, we aim to design a novel spatial interface for image generation tasks to enhance the efficiency of image generation tasks and improve user experience.

\begin{figure*}
\centering
\includegraphics[width=0.7\linewidth]{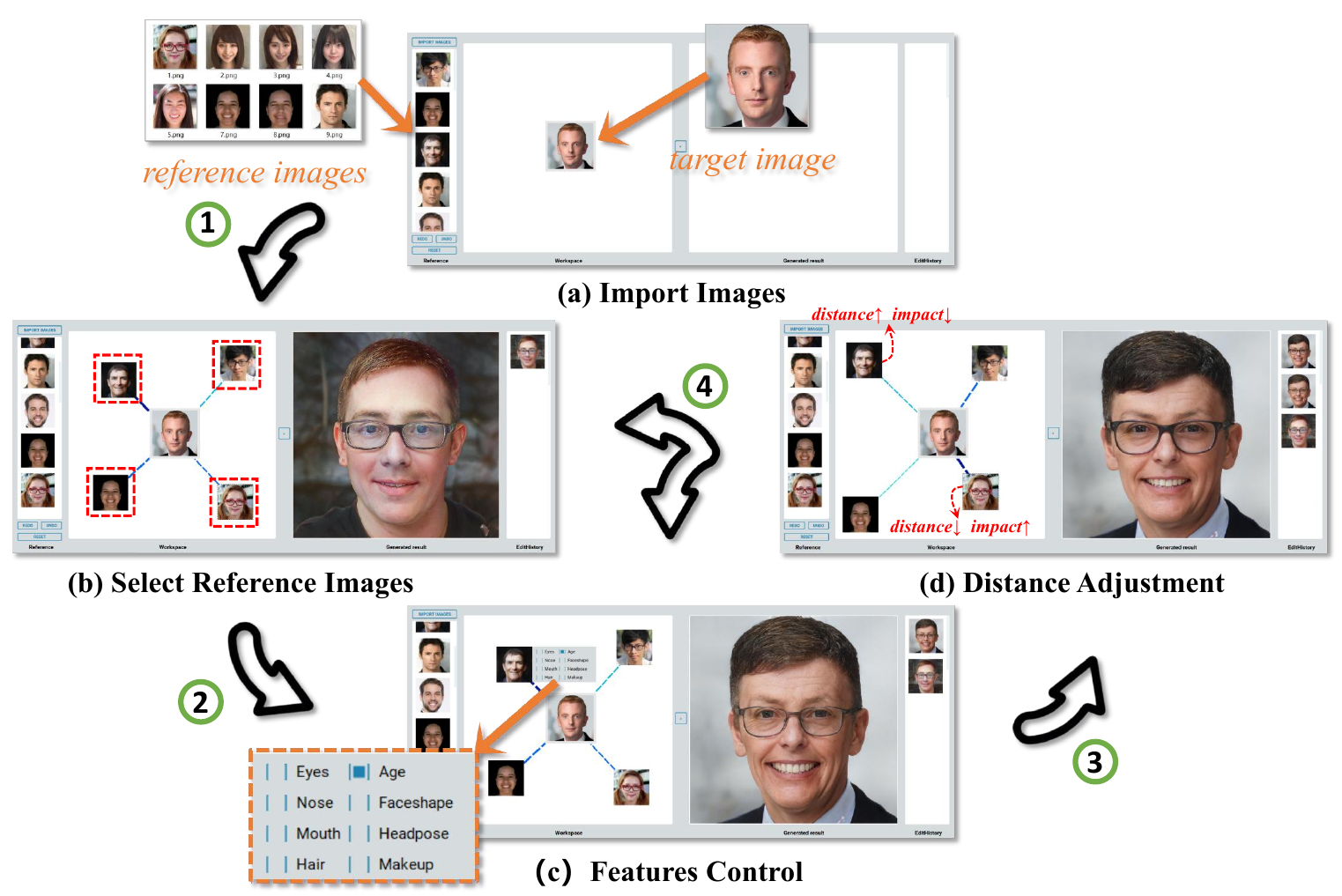}
\caption{\label{fig:workflow}CanvasPic workflow: (a) Importing reference images and the target image. (b) Placing reference images within the spatial layout workspace. (c) Selecting attributes from the reference images. (d) Controlling the generated results by adjusting the distance between the images. Users can then return to step (b) or (c) to continue the image generation process.}
\end{figure*}

\section{TOOL: CanvasPic}

Our CanvasPic is an image generation tool based on 2D spatial layout (see Figure~\ref{fig:workflow}). Users import target and reference images, specify attributes, and manipulate the reference images in the workspace to control the generated result. We have formulated the following design goals for our tool: \textbf{\textit{Goal 1:}} Allow users greater flexibility in generating diverse results. \textbf{\textit{Goal 2:}} Provide users with a simplified and intuitive generative experience through a user-friendly interface layout. \textbf{\textit{Goal 3:}} Offer users easily accessible and efficient image control methods. Additional details are available in the supplementary material.

% Our CanvasPic is an image generation tool based on 2D spatial layout (see Figure~\ref{fig:workflow}). Users import target and reference images, specify attributes, and manipulate the reference images in the workspace to control the generated result. We formulated the following design goals for our tool:

% \textbf{\textit{Goal 1: Allows users greater flexibility to generate diverse results.}} Existing image generation tools limit the representation of attributes, such as slider-based tools. When using sliders to generate specific attributes, users are restricted to choosing from a limited set of preset styles, which limits their ability to achieve more varied results. Transferring rich attribute information from real-world images into the generative process offers a solution to alleviate this constraint.

% \textbf{\textit{Goal 2: Provide users with a simplified and intuitive generative experience through a user-friendly interface layout.}} Existing tools often rely on abstract input methods, such as text input/slider parameters, to control the generated attributes. However, this input method is often challenging to capture the true user needs and lacks intuitiveness. Furthermore, the interfaces of existing tools often lack sufficient scalability and fail to encourage users to explore more in-depth. Therefore, using images as input and integrating them with a two-dimensional interface layout may offer a viable solution to address these challenges.

\subsection{Interface Design and Functions}

\begin{figure*}
\centering
\includegraphics[width=0.8\linewidth]{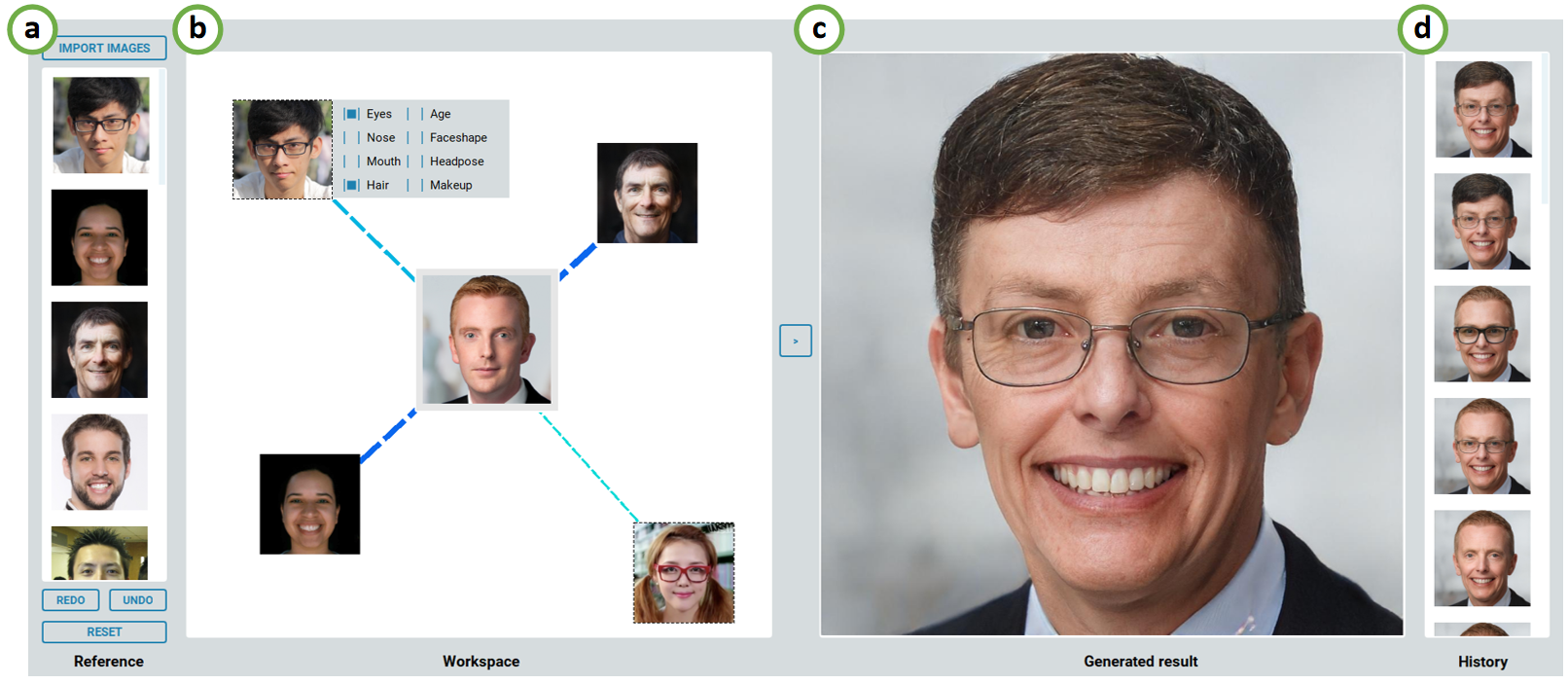}
\caption{\label{fig:teaser}Our interface. The interface comprises four components: a) Reference image bar, b) Spatial layout workspace, c) Results panel, and d) History module.}
\end{figure*}

Our CanvasPic comprises four components: Reference image bar, Spatial layout workspace, Results panel, and History module (see Figure~\ref{fig:teaser}).

\textbf{Reference image bar.} Users can import images from the real world or provided datasets as reference images through the reference image bar \textbf{\textit{(Goals 1 \& 3)}}. The reference image is used to provide attributes for the generated image.

\textbf{Spatial layout workspace.} For CanvasPic, we introduce a novel 2D spatial layout \textbf{\textit{(Goal 2)}}. Users import their desired image (target image) by clicking the central area of the spatial layout workspace. Then, users can add reference images by clicking or dragging them from the reference image bar to the workspace, and connection lines are also added to show their relationships. Right-clicking on a reference image triggers an attribute selection box to allow users to select their desired attributes from the image \textbf{\textit{(Goal 3)}}. 
In our implementation, the attribute selection box is designed to contain eight attributes that are the most critical for facial generation. They are categorized into two groups: local attributes (including eyes, nose, mouth, and hair) and global attributes (including age, faceshape, headpose, and makeup). These attributes in the selection box can be expanded for different tasks.

After that, users can move the reference image to control the influence of selected attributes on the generated results \textbf{\textit{(Goal 3)}}. The closer the reference image is to the target image, the more similar they are, and vice versa. We allow users to redo, undo, or reset their adjustments during the generation process by clicking the appropriate buttons. Additionally, we correspondingly change the color and thickness of the connecting line to further visualize the impact \textbf{\textit{(Goal 2)}}.

\textbf{Results panel.} After users finish their actions in the workspace, they can click the right arrow button to generate a high-resolution image, which will be displayed in the result panel.

\textbf{History module.} We also add a history module that stores all the generated results \textbf{\textit{(Goal 2)}}. Users can click on the generated image to return to the history point.

\subsection{Algorithm}

We utilize pre-trained e4e~\cite{r11e4e} and StyleGAN2~\cite{r11stylegan2} as encoder and generator respectively.

\textbf{Attribute transfer.} Due to the distinct characteristics of local and global attributes, we have explored two algorithms for transferring them individually. For local attributes, we employ a mask preprocessing method inspired by techniques found in StyleMapGAN and r-FACE~\cite{r118, method1}, supplying corresponding masks for each reference image. However, the aforementioned algorithm is not applicable to global attributes. Therefore, we resort to an image code subtraction method inspired by the approach presented in Image2StyleGAN~\cite{method2}.

\textbf{Distance weight calculation.} To realize adjusting the distance between the reference image and the target image impacts the generated result, our tool calculates this distance automatically and assigns an inverse attribute weight to the reference image. Additional algorithmic details can be found in the supplementary material.

\section{Experiments}

Existing generation tools fall into three primary categories: slider-based, text-guided, and sketch-based tools. We chose representative tools from each category: HFGI~\cite{rhfgi}, StyleCLIP~\cite{rstyleclip}, and SketchEdit~\cite{rsketch}. These three tools exemplify the core and widespread forms found within generative tools. HFGI adopts the control mechanism of the commonly used one-dimensional slider tool. StyleCLIP, on the other hand, is among the pioneering tools to employ text-based control for image generation, attracting considerable attention. Meanwhile, SketchEdit represents a foundational form within sketch-based tools. Our experiments focused on local and global attribute image generation tasks, utilizing images from the FFHQ dataset~\cite{r11stylegan}.

\begin{figure}
\centering
\includegraphics[width=0.72\linewidth]{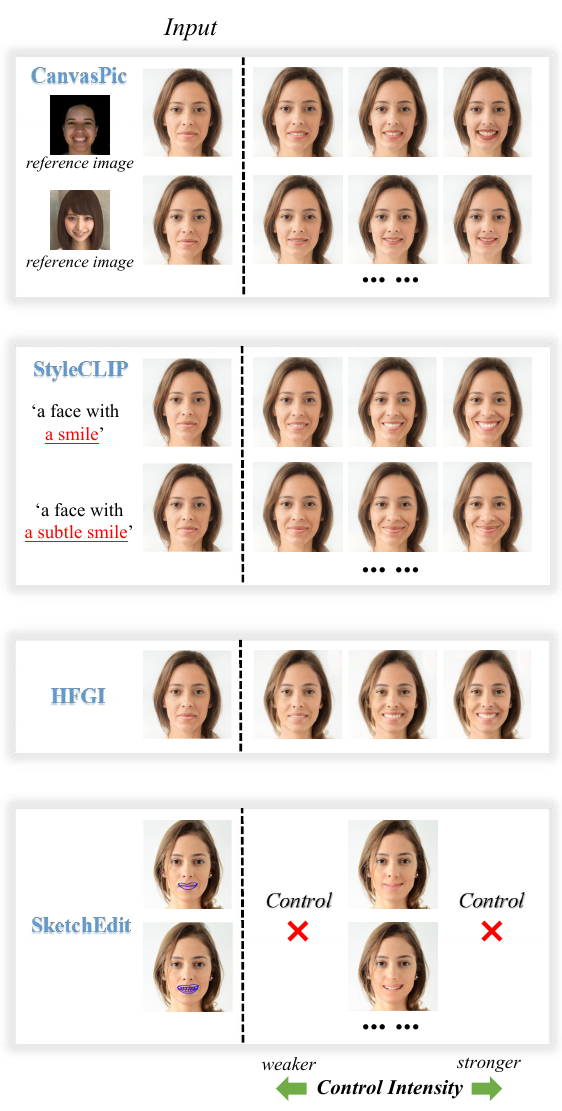}
\caption{\label{fig:img1}Comparison of the results generated by different influence intensities of local attributes. CanvasPic utilizes image distance for intensity adjustment, while StyleCLIP and HFGI use sliders, SketchEdit cannot adjust intensity.}
\end{figure}

\subsection{Local Attribute Generation}

% We employed four different tools to add smiles to portraits (see Figure~\ref{fig:img1}). Among these tools, HFGI, a slider-based tool, revealing limitations associated with such tools for specific tasks. One-dimensional sliders offer restricted control options that limit user control capabilities. Consequently, HFGI can only adjust smiles to a singular style. When users' requirements exceed the predefined slider range, they are unable to achieve the desired generation results.

We used four tools to add smiles to portraits (see Figure~\ref{fig:img1}). HFGI, a slider-based tool, reveals limitations associated with such tools for specific tasks. One-dimensional sliders offer restricted control options that limit user control capabilities. Consequently, HFGI can only adjust smiles to a singular style.

% However, SketchEdit, while also demonstrating a high degree of control freedom, presents some distinct issues. Firstly, images generated by SketchEdit exhibit poor quality, with newly generated regions appearing significantly blurry and lacking proper light and shadow effects. Secondly, the tool is challenging to operate, making it difficult for users without drawing skills to produce desired effects. Finally, there exists a substantial gap between the drawing effects and the generation results, resulting in notable discrepancies between the generated results and users' expectations.

SketchEdit, despite offering considerable control freedom, presents some distinct issues. Firstly, the generated images exhibit poor quality, with blurry regions and lacking proper light and shadow effects. Secondly, the tool is challenging to operate, particularly for users without drawing skills. Lastly, there is a significant gap between the drawing effects and generation results, leading to notable discrepancies between the generated images and users' expectations.

In contrast, the remaining tools provide more flexible generation options by incorporating external reference information, liberating users from predefined constraints. Users can broaden the scope of control by utilizing diverse reference images or input sentences. For example, users can obtain distinct smile effects by introducing reference images with desired expressions or by describing their intended effects through input sentences.

\begin{figure}
\centering
\includegraphics[width=0.72\linewidth]{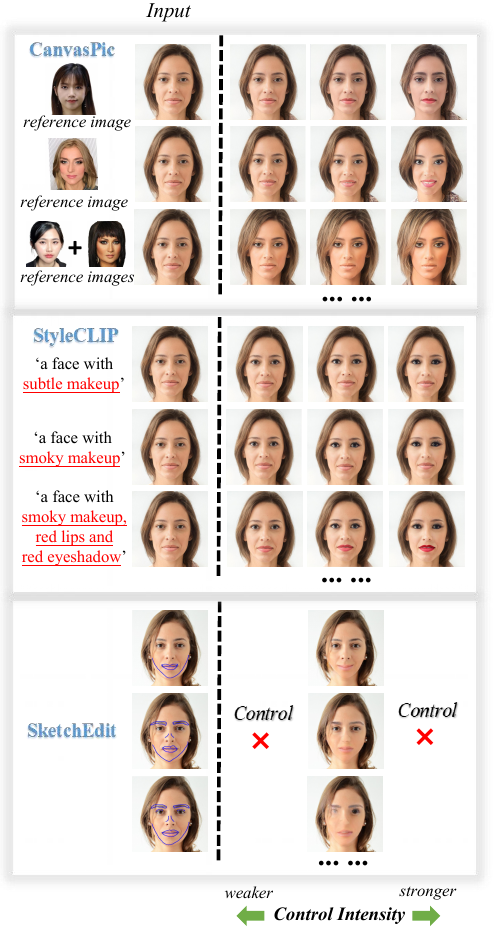}
\caption{\label{fig:img1}Comparison of the results generated by different influence intensities of global attributes.}
\end{figure}

\subsection{Global Attribute Generation}

% We conducted experiments on global attribute generation by applying makeup effects to portraits. For this task, we employed three different tools: our proposed tool, StyleCLIP, and SketchEdit (see Figure~\ref{fig:img1}). We excluded HFGI due to its inability to support makeup generation, a common limitation of slider-based tools. Notably, we identified similar limitations in StyleCLIP and SketchEdit, namely their difficulty in expressing makeup attributes through textual descriptions or drawings. Among them, SketchEdit exhibited poor generation results.

We conducted experiments on global attribute generation, specifically focusing on applying makeup effects to portraits. Three tools were employed for this task: our proposed tool, StyleCLIP, and SketchEdit (see Figure~\ref{fig:img1}). HFGI was excluded due to its inability to support makeup generation. Similar limitations were observed in both StyleCLIP and SketchEdit, particularly when dealing with abstract attributes such as makeup effects, resulting in unsatisfactory results.

% Furthermore, while using StyleCLIP for image generation, we observed a consistent trend of generating images in a similar style, unrelated to the input textual descriptions. This not only increased the challenge for users to achieve expected generation effects but also constrained their expressive abilities. Additionally, achieving specific generation effects using StyleCLIP often required lengthy and detailed textual descriptions. However, this method brought certain challenges. Firstly, StyleCLIP might overlook some details mentioned in complex textual descriptions, such as the red eyeshadow shown in Figure~\ref{fig:img1}. Secondly, providing detailed textual descriptions could be cumbersome. Finally, language may not always accurately convey the intended generation's ideas.

Furthermore, while using StyleCLIP for image generation, we observed remarkably similar generation results regardless of the input text. This hampers users from achieving the desired results and constrains their creativity. Additionally, obtaining specific generation results in StyleCLIP often necessitates detailed textual descriptions, which introduces challenges such as potential oversight of details, as illustrated by the red eyeshadow in Figure~\ref{fig:img1}. Currently, accurately conveying intent through text remains a task that is relatively challenging for users.

% In contrast, we assume that our proposed tool offers a more intuitive and user-friendly way of image generation than existing tools. Users can effortlessly transform images into their desired styles by referring to real-world images. Our tool eliminates the need for complex descriptions and provides users with a direct means to convey their generation's intentions, simplifying the overall experience.

In contrast, we believe our proposed tool provides a more intuitive and user-friendly way of image generation compared to existing tools. Users can easily transform images into desired styles by referencing real-world images. Our tool eliminates the need for complex descriptions and offers users a straightforward means to convey their intentions.

\section{User study}

We conducted a user study to assess the strengths of our proposed tool and its impact on user experience. Following the tests, we conducted semi-structured interviews and distributed questionnaires to gather user feedback.

% We conducted a user study to evaluate: (i) our proposed tool's strengths in different aspects, and (ii) its impact on user experience. 

\begin{table*}[]
\caption{Quantitative results on our questions. The results indicate overall higher satisfaction with CanvasPic among participants.}
\begin{tabular}{ccccccccccccccc}
\toprule[1pt]
\multirow{2}{*}{Comparison tool}&
  \multicolumn{2}{c}{Q1} &
  \multicolumn{2}{c}{Q2} &
  \multicolumn{2}{c}{Q3} &
  \multicolumn{2}{c}{Q4} &
  \multicolumn{2}{c}{Q5} &
  \multicolumn{2}{c}{Q6} &
  \multicolumn{2}{c}{Q7} \\ 
 &
  \multicolumn{1}{c}{\textit{M}} &
  \multicolumn{1}{c}{\textit{SD}} &
  \multicolumn{1}{c}{\textit{M}} &
  \multicolumn{1}{c}{\textit{SD}} &
  \multicolumn{1}{c}{\textit{M}} &
  \multicolumn{1}{c}{\textit{SD}} &
  \multicolumn{1}{c}{\textit{M}} &
  \multicolumn{1}{c}{\textit{SD}} &
  \textit{M} &
  \textit{SD} &
  \textit{M} &
  \textit{SD} &
  \textit{M} &
  \textit{SD} \\ \cline{1-15}
\textit{CanvasPic (Ours)} &
  \textbf{4.50} &
  0.71 &
  \textbf{4.29} &
  0.89 &
  \textbf{3.83} &
  1.11 &
  \textbf{4.21} &
  0.91 &
  \textbf{4.29} &
  0.74 &
  3.96 &
  0.98 &
  \textbf{4.42} &
  0.64 \\
\textit{StyleCLIP} &
  3.21 &
  1.00 &
  3.38 &
  1.07 &
  3.08 &
  1.38 &
  3.33 &
  1.21 &
  3.58 &
  0.91 &
  \textbf{4.04} &
  0.74 &
  3.50 &
  0.96 \\
\textit{HFGI} &
  2.58 &
  1.08 &
  2.83 &
  0.99 &
  3.54 &
  1.04 &
  3.17 &
  1.43 &
  2.83 &
  0.99 &
  3.46 &
  1.26 &
  2.79 &
  1.19 \\
\textit{SketchEdit} &
  2.54 &
  1.29 &
  1.88 &
  1.09 &
  2.54 &
  1.50 &
  2.63 &
  1.25 &
  2.17 &
  1.21 &
  1.92 &
  1.08 &
  2.25 &
  1.23 \\ \bottomrule[1pt]
\end{tabular}
\label{table:tabel1}
\end{table*}

\subsection{Study Design}

\textbf{Tool Selection.} We selected four existing tools: CanvasPic (Ours), HFGI, StyleCLIP, and SketchEdit. To ensure objectivity and anonymity, we assigned distinct code names to each tool. CanvasPic was denoted as $``$Reference Generate$"$ (RG), HFGI as $``$Slider Generate$"$ (SG), StyleCLIP as $``$Text Generate$"$ (TG), and SketchEdit as $``$Sketch Edit$"$ (SE).

\textbf{Participants.} We recruited 24 participants, half male and half female, aged between 20 and 30. All participants have experience using editing software such as Photoshop or Meitu Xiu Xiu, mainly for portrait editing. Experiment durations were set at a maximum of 100 minutes for male participants and 120 minutes for females. After the experiment, each user receives a payment of \$5.

% We set three generation themes:

% \textit{Theme A:} Please use the provided tools to \textbf{design a suitable pair of glasses} for the character in the image.

% \textit{Theme B:} Please use the provided tools to \textbf{add appropriate makeup} for the character in the image.

% \textit{Theme C:} Please use the provided tools to \textbf{change the character to your desired look}.

\textbf{Tasks.} We established three generation themes: designing suitable glasses for the portrait (\textit{Theme A}), adding appropriate makeup to the portrait (\textit{Theme B}), and altering the portrait to your desired look (\textit{Theme C}).

Please note that due to the general lack of experience among males with makeup, therefore male participants are not involved in \textit{Themes B}, while female participants participate in all three themes. Additionally, HFGI and SketchEdit are exclusively designated for \textit{Themes C}. Consequently, female participants use CanvasPic (\textit{Themes A, B, and C}), StyleCLIP (\textit{Themes A, B, and C}), HFGI (\textit{Themes C}), and SketchEdit (\textit{Themes C}) for their tasks, male participants use CanvasPic (\textit{Themes A and C}), StyleCLIP (\textit{Themes A and C}), HFGI (\textit{Themes C}), and SketchEdit (\textit{Themes C}) for their tasks. Each generation task has a maximum duration of 6 minutes.

\textbf{Preparation.} We provided participants with pre-selected images (target images \& reference images) and presented them with three designated image generation themes.

\textit{Target images.} The original images we provide to participants. They use images corresponding to their gender, with males using one specific image and females using another specific image. This differentiation aims to promote participants' sense of identity and minimize gender distinctions, while also facilitating a more diverse range of experimental results.

% This differentiation aims to promote participants' sense of identity and minimize gender distinctions. Additionally, employing different images allows for a more diverse range of experimental results.

\textit{Reference images.} There are a total of 28 reference images, which are only used in CanvasPic, and all participants utilize the same set of reference images.

% \textbf{Interview.} We conducted semi-structured interviews to capture participants' experiences and feedback regarding various tools (see Tabel~\ref{table:tabel1}). To quantify participants' satisfaction with the tools and the generated results, we employed the following evaluation criteria: (1) \textit{Degree of flexibility}: the extent to which users can adjust and control image generation across multiple aspects flexibly; (2) \textit{Ease of implementation}: the level of ease experienced in generating the desired target image; (3) \textit{Effortlessness}: the degree to which the tool reduces effort during image generation; (4) \textit{Intuitiveness}: the degree to which the tool allows users to achieve predictable results, ensuring a close alignment between generated images and user expectations; (5) \textit{Generation effect}: the extent to which external characteristics of subjects are manifested in the final image; (6) \textit{Harmony}: the degree of coherence among characters, scenes, and details in the final image; (7) \textit{Willingness to use}: participants' inclination to utilize this tool for specific use cases or in their daily activities. The Likert five-point scale was employed by the participants to respond to each criterion.

\textbf{Interview.} We conducted semi-structured interviews to capture participants' experiences and feedback (see Tabel~\ref{table:tabel1}). We employed the following evaluation criteria: (Q1) \textit{Degree of flexibility}; (Q2) \textit{Ease of implementation}; (Q3) \textit{Effortlessness}; (Q4) \textit{Intuitiveness}; (Q5) \textit{Generation effect}; (Q6) \textit{Harmony}; (Q7) \textit{Willingness to use}. The Likert five-point scale was employed by the participants to respond to each criterion.

\subsection{Quantitative Results}
In this section, we use the abbreviation P to designate participants. For example, we represent the first participant labeled as P1. All statistical tests are conducted at a significance level ($\alpha$) of 0.05.

We used the Friedman test to analyze our data. For Q1, twenty-one participants regarded CanvasPic as having the highest degree of flexibility (\textit{M}=4.50, \textit{SD}=0.71)  (\textit{s} = 36.38, \textit{p} < .05) (Tabel~\ref{table:tabel1}: Q1). Similarly, nineteen participants found CanvasPic to be highly accessible for implementation (\textit{M}=4.29, \textit{SD}=0.89) (\textit{s} = 43.90, \textit{p} < .05) (Tabel~\ref{table:tabel1}: Q2). Regarding Q3, sixteen participants perceived the tool to exhibit strong effortlessness (\textit{M}=3.83, \textit{SD}=1.11) (\textit{s} = 12.89, \textit{p} < .05) (Tabel~\ref{table:tabel1}: Q3). Among the twenty-four participants, eighteen reported that CanvasPic demonstrated excellent intuitiveness compared to the other three tools (\textit{M}=4.21, \textit{SD}=0.91) (\textit{s} = 19.18, \textit{p} < .05) (Tabel~\ref{table:tabel1}: Q4). Concerning Q5, twenty participants expressed satisfaction with CanvasPic's generative performance in comparison to the other three tools (\textit{M}=4.29, \textit{SD}=0.74) (\textit{s} = 39.75, \textit{p} < .05) (Tabel~\ref{table:tabel1}: Q5). Sixteen participants believed that CanvasPic's generated results possessed the highest level of harmony (\textit{M}=3.96, \textit{SD}=0.98) (\textit{s} = 39.02, \textit{p} < .05) (Table~\ref{table:tabel1}: Q6), slightly inferior to StyleCLIP, which eighteen participants also found impressive (\textit{M}=4.04, \textit{SD}=0.74) (Table~\ref{table:tabel1}: Q6). Furthermore, twenty participants exhibited a stronger intention to use CanvasPic (\textit{M}=4.42, \textit{SD}=0.64) (\textit{s} = 37.61, \textit{p} < .05) (Tabel~\ref{table:tabel1}: Q7). In summary, CanvasPic yielded satisfactory results in terms of flexibility, ease of implementation, intuitiveness, and generative performance for the participants.

\subsection{Qualitative Results}

After each task, we conducted semi-structured interviews to learn more about the participants' experiences.

\textbf{Participants showed a clear preference for tools with high degrees of freedom and flexible control options.} In the interviews, three-quarters of participants expressed a preference for tools with a high degree of freedom, and twenty-one believed that our tool offered the greatest level of freedom. The reasons include:

\textit{Facilitate goal achievement.} P6 mentioned, $``$The RG tool offers me more control options, helping me achieve results that better align with the theme.$"$
\textit{Nurture creativity.} P4 stated, $``$High-degree-of-freedom tools showcase more possibilities, making me more interested in generating images and trying various combinations.$"$
\textit{Improve user experience.} Specifically, six participants noted that high-degree-of-freedom tools provided smoother and more seamless control, significantly enhancing their usage experience.

\textbf{Participants showed a clear preference for intuitive modes of image generation.} The majority of participants preferred intuitive image generation tools, with eighteen finding our tool to be the most intuitive. The reasons include:

\textit{Generating images from real-world references is intuitively effective.} P14 noted, $``$When dealing with thematic tasks, I have a certain idea of the generated image. I can use reference images to find matching attributes to achieve the desired result.$"$ However, existing methods often use abstract controls such as preset sliders or text input, leading to significant discrepancies in results and user expectations.
\textit{The two-dimensional layout enhances intuitiveness.} P11 mentioned, $``$When performing multi-attribute control, the two-dimensional layout of the RG tool effectively indicates the attributes represented by reference images, allowing me to predict the generated results to some extent.$"$

% \textbf{Participants exhibited a clear inclination toward employing simple yet effective control methods.} During the interviews, 14 participants expressed that the proposed tool (RG) provides a simple yet effective control mechanism. For instance, P3 mentioned, $``$Sketching and describing what you want in words is limited and cumbersome, but using images can effectively avoid this issue.$"$ Furthermore, participants also demonstrated a clear preference for concise control methods that optimize the user experience. For example, P2 pointed out, $``$If I can only use the mouse for control, I wouldn't want to use the keyboard.$"$ 

\section{DISCUSSION and future work}

\subsection{Human-AI Collaborative Image Creation}

% AI-driven image generation tools hold immense potential in fostering collaboration between humans and artificial intelligence, creating new opportunities for synergy by merging human intuition and creativity with AI's computational prowess. For instance, P17 mentioned, $``$While engaging in the task of designing glasses, I found myself creating styles I had never seen before. It's interesting, and I believe it could hold significant value in the design field.$"$ This collaborative method not only yields exceptional design results but also broadens creators' perspectives, infusing the field of image generation with heightened creativity and uniqueness.

AI-driven image generation tools hold immense potential in fostering collaboration between humans and artificial intelligence, creating new opportunities for synergy by merging human intuition and creativity with AI's computational prowess. For instance, P17 mentioned, $``$While engaging in the task of designing glasses, I found myself creating styles I had never seen before. It's interesting, and I believe it could hold significant value in the design field.$"$ 

% \textbf{Precisely and freely controlling image generation.} In the realm of modern creativity and design, AI-driven interfaces have exerted a significant impact. Interactive tools enable users to more manageably and intuitively control the GAN-based image generation process. For instance, P2 stated, $``$The RG tool has excellent intuitiveness and control, making it easier for me to achieve desired results.$"$ This perspective is consistent with the experimental findings indicating that CanvasPic offers users with better ways to explore their creative visions. By providing a more controllable image generation process, our tool empowers users with greater creative freedom and expressive capacity, facilitating the transformation of ideas into reality and unlocking broader possibilities for creators in diverse fields.

\textbf{Precisely and freely controlling image generation.} In the realm of modern creativity and design, AI-driven interfaces have exerted a significant impact. Interactive tools enable users to more manageably and intuitively control the GAN-based image generation process. For instance, P2 stated, $``$The RG tool has excellent intuitiveness and control, making it easier for me to achieve desired results.$"$ This aligns with the experimental results, indicating that CanvasPic provides users with greater creative freedom and expressive capacity, opening up broader possibilities for creators.

\textbf{Impact of diverse input on user experience.} Existing input methods can be broadly categorized into text and sketch inputs. In this paper, text input allows a certain degree of image control but exhibits limitations in fine adjustments and intuitiveness, sketching may require a certain level of artistic skill. Our proposed method, generating images using attributes from reference images, addresses limitations in both text and sketch methods. It avoids the ambiguity of text input and the skill requirements of sketching. Additionally, our method simplifies the input process, making it easier for users to achieve their goals. Experimental results confirm that image input outperforms text and sketch input.

% \textbf{AI can serve as a medium for communication and exchange of opinions among users.} \cite{raic1,rwetoon,raic2} also introduces this point. P4 mentioned, $``$I believe that the RG tool could enhance task completion through communication. If clients display the images they want to generate and provide feedback, they can present the desired results as reference images, rather than having me guess based on vague text.$"$ 
% This implies that such tools can not only better meet user communication needs but also provide clear guidance, reducing misunderstandings and unnecessary revisions among users.

% \subsection{Extension and Limitations}
% \subsection{Limitations}
\subsection{Future Work}

% We further extend the proposed 2D layout to generate images sequentially (see Figure~\ref{fig:ex}). This feature enables users to generate new images by leveraging their previously generated results. A hierarchical tree-like structure serves as a conduit for transmitting these relationships, facilitating deeper exploration of interconnections in image generation.

% Firstly, our current implementation only works with close-up facial shots, since facial generation is one of the most prominent applications and is rich in practical information. We plan to expand its versatility by incorporating different data and models in the future. Secondly, we use masks to control local attributes. The accuracy of masks affects the quality of generated results, especially for complex shapes. We will focus on solving this problem in our further work.

% The current version of our tool is designed for close-up facial shots. In future updates, we plan to enhance its generalization by training it on more diverse datasets, extending its application to various scenarios such as different parts of the human body, animals, and natural landscapes. Additionally, we aim to explore the strengths of different models in specific domains and integrate them appropriately to broaden the functionality of our tool. Lastly, our method involves using masks to control local attributes. However, the accuracy of masks affects the quality of generated results. In our future work, we intend to incorporate lightweight object detection algorithms to improve the automation and accuracy of our method.

Our current tool is specifically designed for close-up facial shots. In future updates, we aim to improve its versatility by training it on more diverse datasets, enabling it to handle various scenarios, including different parts of the human body, animals, and natural landscapes. We also plan to explore the strengths of different models in specific domains and integrate them to enhance the overall functionality of our tool. Currently, our method involves using masks to control local attributes, and the accuracy of these masks influences the quality of the generated results. In our upcoming work, we intend to incorporate lightweight object detection algorithms to automate and improve the accuracy of our method.

% Firstly, our current tool is limited to close-up facial shots. In future releases, we plan to enhance its generalization by training on more diverse datasets, extending applicability to various scenarios such as different parts of the human body, animals, natural landscapes, and buildings. Secondly, we aim to investigate the strengths of different models in specific domains and integrate them appropriately to broaden the functionality of our tool. Lastly, our approach involves using masks to control local attributes. However, the accuracy of masks affects the quality of generated results. In our future work, we intend to incorporate lightweight object detection algorithms to improve the automation and accuracy of our method.

\section{CONCLUSION}

% We introduce a novel GAN-based image generation tool, CanvasPic, aimed at providing users with enhanced freedom and intuitiveness in their creative experience. By incorporating a 2D spatial layout design, users can interact with elements within the layout, effortlessly adjust image attributes, and delve into exploring diverse potential results. We conducted a user study, revealing that CanvasPic outperforms existing tools in terms of freedom, intuitiveness, user experience, and generated results. We hope that the results of this study provide insights into the future development of image generation tools.

We introduce a novel GAN-based image generation tool, CanvasPic, aimed at providing users with enhanced freedom and intuitiveness in their creative experience. By incorporating a 2D spatial layout design, users can interact with elements within the layout, effortlessly adjust image attributes, and delve into exploring diverse potential results. Our user study indicates that CanvasPic outperforms existing tools in freedom, intuitiveness, user experience, and generated results. We hope that the results of this study provide insights into the future development of image generation tools.

\begin{acks}
This work was supported by Jilin University (Grant No.419021422B08), JST ACT-X Grant Number JP-MJAX21AG and JST CREST Grant Number JP- MJCR17A1, Japan.
\end{acks}

\bibliographystyle{ACM-Reference-Format}
\bibliography{sample-base}

\appendix

\end{document}